\documentclass[
journal=jacsat, 
manuscript=article]{achemso}

\usepackage[version=3]{mhchem} 

\author{Bodo Zibrowius}
\email{zibrowius@mpi-muelheim.mpg.de}
\author{Michael Felderhoff}
\email{felderhoff@mpi-muelheim.mpg.de}
\affiliation
{Max-Planck-Institut f\"ur Kohlenforschung, Kaiser-Wilhelm-Platz 1, 45470 M\"ulheim an der Ruhr, Germany}

\title[\texttt{achemso} demonstration]
{On the preparation and NMR spectroscopic characterization of potassium aluminium tetrahydride KAlH$_{4}$}

\begin{document}

\begin{abstract}
Potassium aluminium tetrahydride \ce{KAlH4} of high phase purity (space group \textit{Pnma} (62)) was synthesized via a mechanochemical route. The thus obtained material was studied by $^{27}$Al and $^{39}$K MAS NMR spectroscopy. For both nuclei precise data for the isotropic chemical shift and the quadrupole coupling at $T=295$\,K were derived ($^{27}$Al: $\delta_{iso}=(107.6\pm 0.2)$\,ppm,  $C_Q = (1.29\pm0.02)$\,MHz and $\eta = 0.64\pm0.02$; $^{39}$K: $\delta_{iso}=(6.1\pm0.2)$\,ppm, $C_Q = (0.562\pm0.005)$\,MHz and $\eta = 0.74\pm0.02$).  The straightforward NMR spectroscopic approach applied here should also work for other complex aluminium hydrides and for many other materials containing half-integer nuclei experiencing small to medium-sized quadrupole couplings.
\end{abstract}

\section{Introduction}

In the quest for materials for reversible hydrogen storage, complex aluminium hydrides have attracted much scientific interest in the last two decades.\cite{Schueth04, Orimo07, Li13} Because of its excellent dehydrogenation reversibility under the influence of a catalyst at moderate conditions,\cite{Bogdanovic97}        sodium aluminium tetrahydride \ce{NaAlH4} is certainly one of the most studied potential hydrogen storage materials. Besides other methods, X-ray\cite{Bogdanovic03, Brinks04, Hauback08, Humphries14}  and neutron diffraction,\cite{Hauback03, Brinks04, Hauback08, Humphries14} Raman scattering\cite{Ross04, Majzoub05, Yukawa07} and X-ray absorption\cite{Balde07, Leon07} and nuclear magnetic resonance (NMR) spectroscopy \cite{Bogdanovic03, Zhang09, Verkuijlen09, Verkuijlen11} have been applied to study \ce{NaAlH4} and its decomposition in great detail. Nevertheless, some important questions concerning the mechanism of hydrogen release and uptake remain open.\cite{Frankcombe12, Kocabas15} 

The number of papers devoted to the characterization of the next heavier alkali aluminium hydride KAlH$_{4}$ is rather limited. Morioka \textit{et al.}\cite{Morioka03} have shown that potassium aluminium hydride can reversibly be dehydrogenated and rehydrogenated in the temperature range from 523 to 613 K without any catalysts. The decomposition of  KAlH$_{4}$ into potassium hydride and aluminium metal releases only 4.3\,wt.\% hydrogen compared to 5.6\,wt.\% for the corresponding decomposition of NaAlH$_{4}$. So, KAlH$_{4}$ is of course a rather unlikely candidate for a widespread application as hydrogen storage material.  Nevertheless, the elucidation of structural details and mechanisms of dehydrogenation and rehydrogenation of KAlH$_{4}$ and other alkali aluminium hydrides, interesting in its own right, may help to better understand the more promising NaAlH$_{4}$ hydrogen storage system. Indeed, studies of the decomposition of KAlH$_{4}$ have revealed the formation of an unknown intermediate either instead of\cite{Mamatha06} or alongside with\cite{Ares09} the expected formation of the hexahydride K$_{3}$AlH$_{6}$. In an \textit{in-situ} powder X-ray diffraction study using synchrotron radiation, three different intermediates with hitherto unknown structures have been found during the decomposition and formation of KAlH$_{4}$.\cite{Arnbjerg12} More recently, Ares \textit{et al.}\cite{Ares16} have detected only \ce{K3AlH6} as intermediate in both the mechanochemical synthesis and the subsequent decomposition of \ce{KAlH4}. These at least partially contradictory findings might indicate that the release and uptake reactions in alkali aluminium hydrides are more complex than hitherto thought and might also depend on the experimental conditions used.

There are at least two straightforward preparation methods for KAlH$_{4}$. The first is a direct synthesis reported by Ashby \textit{et al.}\cite{Ashby63} starting from metallic potassium (or potassium hydride) and metallic aluminium under hydrogen pressure at elevated temperatures in diglyme. After precipitation with a nonpolar solvent, filtration and careful drying, a pure KAlH$_{4}$ material can be obtained. Alternatively, the direct hydrogenation can be performed under mechanochemical conditions. \cite{Pukazhselvan15, Ares16} Another, less demanding preparation method for KAlH$_{4}$ is the salt metathesis reaction starting from a potassium halide and NaAlH$_{4}$ or LiAlH$_{4}$ under mechanochemical conditions.\cite{Mamatha06} To remove the produced alkali halide, a subsequent treatment with diglyme is necessary. The pure KAlH$_{4}$ can once again be obtained after precipitation, filtration and drying.

To the best of our knowledge, there are only four reports on NMR studies of potassium aluminium tetrahydride. Tarasov \textit{et al.}\cite{Tarasov00} studied KAlH$_{4}$ by using $^{27}$Al and $^{39}$K NMR on stationary (non-spinning) samples. They reported chemical shift and quadrupole coupling data for both nuclei. Surprisingly, the authors found three different aluminium species with distinctly different quadrupole couplings. They interpreted their findings in terms of a static orientational disorder of the $[$AlH$_{4}]^{-}$ anions. According to the widely accepted structure solutions for \ce{KAlH4}\cite{Chung04} and \ce{KAlD4},\cite{Hauback05} all aluminium atoms are equivalent. Later on, Tarasov \textit{et al.}\cite{Tarasov01} studied the thermal decomposition of KAlH$_{4}$ by the same methods. They identified the decomposition products potassium hydride KH and aluminium metal and the intermediate hexahydride K$_{3}$AlH$_{6}$. 

The third, more recently published paper by Sorte \textit{et al.}\cite{Sorte14} is mainly devoted to the study of the dynamics of the $[$AlH$_{4}]^{-}$ anions. To this end, $^{1}$H and $^{27}$Al NMR measurements on a stationary sample were performed in a wide temperature range. The experimental results with respect to the quadrupole coupling do not really fit those obtained earlier. The $^{27}$Al NMR line measured for the stationary sample at room temperature is broad, but the singularities typical of quadrupole coupling are missing. This holds also for the sideband pattern observed in the $^{27}$Al MAS NMR spectrum. The authors assume ``that the quadrupolar singularities were broadened by some disorder, resulting in a distribution of the parameters'' which are describing the quadrupole coupling. For the additionally given $^{39}$K NMR spectrum, the disagreement with the earlier work\cite{Tarasov00} is even more evident. Sorte \textit{et al.}\cite{Sorte14} finally arrive at the rather unsatisfactory conclusion ``that there may be other crystal structures present in our KAlH$_{4}$ or that used in the previous study.'' In the fourth, above already mentioned paper,  Ares \textit{et al.}\cite{Ares16} used
$^{27}$Al and $^{1}$H MAS NMR spectroscopy to characterize their KAlH$_{4}$ samples obtained via direct hydrogenation under ball milling conditions. Although the spectral resolution achieved in the $^{27}$Al MAS NMR spectra is way better than in the earlier papers, the authors have to admit that the ''lack of sharp feature in the first-order quadrupolar spinning-sideband pattern suggests that structural disorder exists.'' Since materials produced by ball milling usually consist of very small particles with a large number of defects this finding is not surprising.

Recent papers\cite{Bogdanovic03, Zhang09, Verkuijlen09} on the sodium aluminium hydrides NaAlH$_{4}$ and Na$_{3}$AlH$_{6}$ have shown that the NMR parameters, which mirror the site geometry, can reliably be determined for both the aluminium atom and the alkali cation if well-defined alkali aluminium hydride samples are used. In our very recent study on CsAlH$_{4}$ \cite{Krech14} we were able to unambiguously distinguish two different polymorphs present in the freshly precipitated hydride on the basis of well resolved $^{27}$Al and $^{133}$Cs MAS NMR spectra. 

The aim of the present paper is to demonstrate that KAlH$_{4}$ with the previously described structure \cite{Hauback05} can readily be synthesized following the established metathesis procedure. The thus obtained material exhibits a sufficient crystallinity and phase purity and allows the NMR chemical shift and quadrupole coupling parameters to be determined with high accuracy. The precise determination of these parameters combined with state-of-the-art DFT calculations offers an alternative approach to high-quality structures for polycrystalline materials. This NMR crystallographic approach is particularly helpful when diffraction-based methods are hampered by crystal twinning and stacking faults.\cite{Perras12} 

To ease the understanding of the procedure used for the extraction of the relevant parameters, we give a brief summary of important peculiarities of solid-state NMR spectroscopy of quadrupolar nuclei. The procedure outlined for $^{27}$Al can be considered as a blueprint for NMR studies of materials containing half-integer quadrupolar nuclei with small to medium values of the coupling constant.

\section{Experimental}

\subsection{Synthesis of KAlH$_{4}$}

All procedures were performed under argon as protective gas atmosphere and with carefully dried solvents. 3.00\,g (40.2\,mmol) KCl and 2.15\,g (40.2\,mmol) of NaAlH$_{4}$ were mechanochemically treated in a planetary ball mill for 3\,h (Pulverisette P7, 500 rpm, hardened steel milling vial 45\,ml with 7 balls 13.5\,g each, 15\,min milling, 10\,min break after each milling period, 12 repetitions). The resulting powder was suspended in 25\,ml diglyme and stirred for 12\,h. After filtration, KAlH$_{4}$ was precipitated through the addition of toluene and filtered off. The carefully dried KAlH$_{4}$ contained small amounts of unreacted NaAlH$_{4}$ observable in the X-ray diffraction pattern. Pure KAlH$_{4}$ was obtained after treatment with 20\,ml tetrahydrofurane for 12\,h, filtration and drying in vacuum. 

\subsection{X-ray diffraction}

The X-ray diffraction (XRD) measurements were carried out in Debye-Scherrer transmission geometry on a STOE STADI P diffractometer using Cu K$\alpha_{1}$ ($\lambda$ = 1.5460\,\AA) radiation. To avoid any contact with air, sealed 0.5\,mm diameter glass capillaries were used. The data were collected with a linear position sensitive detector fabricated by Stoe in the 15--80$^\circ$ 2$\Theta$ range.

\subsection{Solid-state NMR spectroscopy}

The NMR spectra were recorded on a Bruker Avance\,III\,HD 500WB spectrometer using double-bearing MAS probes (DVT BL4) at resonance frequencies of 130.3\,MHz and 23.3\,MHz for $^{27}$Al and $^{39}$K, respectively. The magic angle was adjusted by maximizing the rotational echoes of the $^{23}$Na resonance of solid \ce{NaNO3}. 

For the $^{27}$Al MAS~NMR spectra, single $\pi$/12 pulses ($t_p$ = 0.6\,$\mu$s) were applied at a repetition time of 2\,s (2,000--16,000 scans) and spinning frequencies ($\nu_{MAS}$) between 1.4 and 8.0\,kHz.  High-power proton decoupling (SPINAL-64) was used for all spectra shown in this paper, although its influence on the spectral resolution was hardly noticeable at spinning frequencies above 5\,kHz. The $^{27}$Al chemical shifts were referenced relative to an external 1.0 M aqueous solution of aluminium nitrate. The same solution was used for determining the flip-angle. 

For the $^{39}$K MAS NMR spectra, single $\pi$/8 pulses ($t_p$ = 3.0\,$\mu$s) were applied at a repetition time of 6\,s (4,000--24,000 scans) and spinning frequencies between 4.0 and 6.0\,kHz.
The $^{39}$K chemical shifts are given with respect to 1.0\,M aqueous solution of potassium chloride using solid potassium chloride ($\nu_{MAS}$ = 6\,kHz) with its sharp resonance at $\delta$= 47.8\,ppm\cite{Hayashi90} as secondary standard. Solid potassium chloride was also used for determining the flip-angle. 

Both the spinning-sideband patterns of the $^{27}$Al MAS~NMR spectra and the centrebands of the $^{39}$K MAS NMR spectra were simulated by using the solids lineshape analysis module implemented in the TopSpin\texttrademark\ 3.2 NMR software package from Bruker BioSpin GmbH. Throughout the paper any effect of chemical shift anisotropy on the $^{27}$Al NMR lineshape is neglected.

\begin{figure}
  \includegraphics [width=15cm]{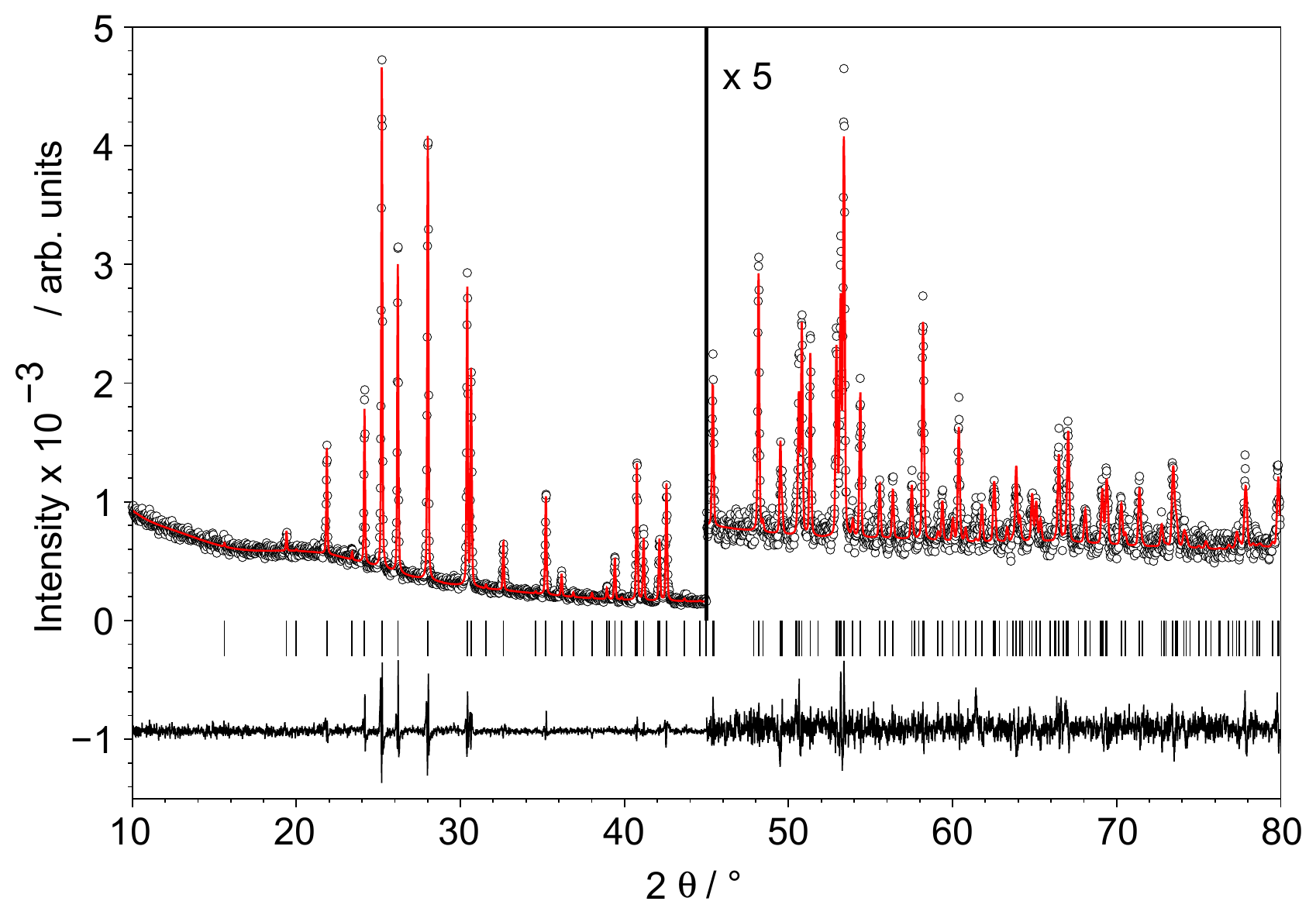}
  \caption{X-ray powder diffraction pattern of \ce{KAlH4}: black circles for the measured intensities ($I_0$), red line for the calculated intensities ($I_c$, \textit{cf.} text) and black line (bottom trace) for the difference ($I_0-I_c$). The black tick-marks represent the positions of the Bragg reflections.}
  \label{fgr:KAlH4_XRD}
\end{figure}

\section{Results and discussion}
\subsection{Structure}

Fig.~\ref{fgr:KAlH4_XRD} shows the X-ray powder diffraction pattern  measured for our \ce{KAlH4} sample in comparison with a calculated diffraction pattern based on the structural data for \ce{KAlD4}.\cite{Hauback05} To account for the difference between \ce{KAlH4} and \ce{KAlD4}, the lattice parameters (in \AA) were slightly adjusted: $a=$ 8.8475\,(8.8514), $b=$ 5.81428\,(5.8119) and $c=$ 7.3448\,(7.3457). The values in parentheses are those reported by Hauback \textit{et al.}    \cite{Hauback05} for \ce{KAlD4} at $T=$ 295\,K. The agreement between the calculated and the observed pattern shows that the material under study is structurally identical to the previously reported orthorhombic potassium aluminium tetrahydride \ce{KAlH4}.\cite{Bastide87, Morioka03, Chung04, Hauback05} No crystalline by-product can be identified in the powder pattern in Fig.~\ref{fgr:KAlH4_XRD}. According to the structure solutions for \ce{KAlH4} in space group \textit{Pnma}(62)\cite{Hauback05} or \textit{Pbnm}(62),\cite{Chung04} both the aluminium and the potassium atoms as well as half of the hydrogen atoms  are located in Wyckoff position 4c, i.e., they sit on mirror planes.  

\subsection*{$^{27}$Al MAS NMR spectroscopy}
\begin{figure}
  \includegraphics [width=10cm]{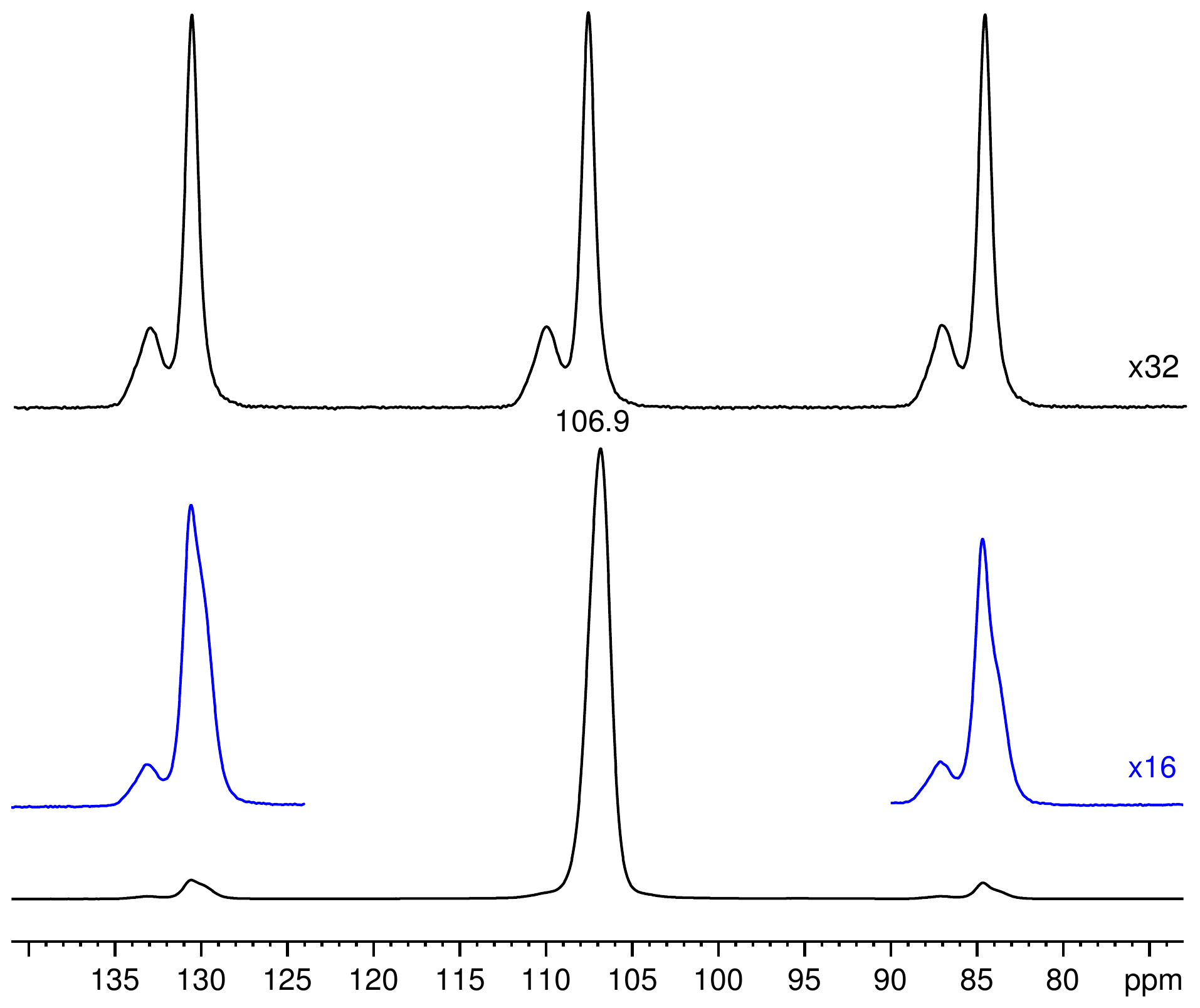}
  \caption{$^{27}$Al MAS NMR spectrum of KAlH$_{4}$ measured at the rather low spinning frequency $\nu_{MAS}$ = 3 kHz (bottom trace). The top trace depicts the enlarged spinning sidebands $+$3 to $+$5 shifted by $-$4$\nu_{MAS}$. The two blue inserts show the enlarged first ($\pm$1) spinning sidebands.}
  \label{fgr:KAlH4_27Al_centre}
\end{figure}

Different regions of a $^{27}$Al MAS NMR spectrum of the above structurally characterized KAlH$_{4}$ sample are depicted in Fig.~\ref{fgr:KAlH4_27Al_centre}. The bottom trace shows the centreband region with an almost symmetrical line at 106.9\,ppm. This position of the resonance line agrees with those reported in the previous studies of KAlH$_{4}$\cite{Tarasov00,Sorte14} and fits into the range generally obtained for isolated $[$AlH$_{4}]^{-}$ units in alkali aluminium hydrides.\cite{Tarasov08} The linewidth of about 190\,Hz (full width at half height, $FWHH$) is comparable with that measured for the isostructural orthorhombic CsAlH$_{4}$ under the same experimental conditions.\cite{Krech14} 
It should be noted that spinning the sample at about 3 kHz results in the best spectral resolution. A further increase in the spinning speed leads to a significant broadening and a high-field shift of the centreband. This effect can easily be explained by a non-uniform increase in temperature over the length of the sample brought about by frictional heating.\cite{Brus00,Antonijevic05} At $\nu_{MAS}$ = 8\,kHz, a centreband with a width of about 230\,Hz is observed at 106.8\,ppm. Although no indication of crystalline by-products was found in the XRD powder pattern of the sample after the treatment with tetrahydrofurane, a close inspection of the spectrum in Fig.~\ref{fgr:KAlH4_27Al_centre}  reveals the presence of traces of both \ce{NaAlH4} and \ce{Na3AlH6} (see ESI,~Fig.~S1).

Since $^{27}$Al is a nucleus with a spin $I = \tfrac {5}{2}$, the NMR spectrum consists not only of the central transition $(-\tfrac{1}{2}\leftrightarrow+\tfrac{1}{2})$, but comprises also the pair of inner satellite transitions $(+\tfrac{1}{2}\leftrightarrow+\tfrac{3}{2})$ and $(-\tfrac{3}{2}\leftrightarrow-\tfrac{1}{2})$   and the pair of outer satellite transitions $(+\tfrac{3}{2}\leftrightarrow+\tfrac{5}{2})$ and  $(-\tfrac{5}{2}\leftrightarrow-\tfrac{3}{2})$.\cite{Freude00, Man12} These pairs of satellite transitions are symmetric, i.e., the lineshape of a satellite transition $(-m\leftrightarrow -m+1)$ is a  mirror image of the lineshape of the transition  $(m-1\leftrightarrow m)$. The quadrupole interaction does not only cause several, usually broad, resonance lines per nucleus, but also shifts the positions of the resonance lines with respect to those given by the Larmor equation. This additional shift of the resonance lines occurs with and without sample spinning. Hence, as distinct from the case of nuclei with $I = \tfrac {1}{2}$, the position of the centreband is not identical with the isotropic chemical shift $\delta_{iso}$. Instead, the position of the centre of gravity of the resonance line $\delta_{cg}$ of the $(m-1\leftrightarrow m)$ transition of a quadrupolar nucleus differs from $\delta_{iso}$ by the so-called quadrupole-induced shift $\delta_{qis}(m)$:

\begin{equation}
  \delta_{cg}(m) = \delta_{iso} + \delta_{qis}(m).
  \label{eq:a}
\end{equation}
The quadrupole-induced shift $\delta_{qis}(m)$ depends on the spin quantum numbers $I$ and $m$, the quadrupole coupling constant $C_Q$ and the asymmetry parameter $\eta$ of the nucleus under study and the magnetic field at which the spectrum is measured.\cite{Samoson85} The quadrupole coupling constant is defined as 
\begin{equation}
  C_Q = \frac{e^2qQ}{h},
  \label{eq:z1}
\end{equation}
where $eq=V_{zz}$ is the $zz$-component of the electric-field gradient tensor, $e$ is the elementary charge, and $Q$ is the quadrupole moment of the nucleus.\cite{FreudeHaase} With the usual convention for the principal tensor components, $|V_{zz}| \geq |V_{yy}| \geq |V_{xx}|$, the asymmetry parameter is definded as
\begin{equation}
  \eta = \frac{V_{xx}-V_{yy}}{V_{zz}}.
  \label{eq:z2}
\end{equation}
As shown by Samoson,\cite{Samoson85} the quadrupole-induced shift of a transition $(m-1\leftrightarrow m)$ of a nucleus with spin $I$ can be calculated by the following formula:
\begin{equation}
  \delta_{qis}(m)=-\frac{3}{40} \frac{C_Q^2}{\nu_L^2}\frac{I(I+1)-9m(m-1)-3}{I^{2}(2I-1)^{2}}\left(1+\frac{\eta^2}{3}\right).
  \label{eq:z}
\end{equation}
Here, $\nu_L$ is the Larmor frequency divided by 2$\pi$. 

Although the effects of the quadrupole interaction on shape and position of NMR lines are well  known to experts in the field, line positions observed in NMR spectra of quadupolar nuclei are very often erroneously reported as chemical shifts. For the central transition, the quadrupole-induced shift is always negative, i.e., the resonance line is shifted to higher field. By determining $\delta_{cg}(m)$ for two different values of $m$, the isotropic chemical shift $\delta_{iso}$ can be experimentally obtained in a reliable way. In particular, for a nucleus with $I = \tfrac {5}{2}$ the quadrupole-induced shift of the inner satellite transitions is given by $\delta_{qis}(\tfrac {3}{2}) = -\delta_{qis}(\tfrac {1}{2})/8$.\cite{Samoson85} Hence, the isotropic chemical shift can easily be calculated from the following relation:
\begin{equation}
  \delta_{iso} = \delta_{cg}(\tfrac {3}{2}) - \left(\delta_{cg}(\tfrac {3}{2}) - \delta_{cg}(\tfrac {1}{2})\right)/9.
   \label{eq:b}
\end{equation}
If the quadrupole coupling is small, $\delta_{iso}$ is very close to the centre of gravity of the inner satellite transitions. 

For the sample under study, the centreband and the first spinning sidebands (enlarged in the blue insets above the bottom trace of Fig.~\ref{fgr:KAlH4_27Al_centre}) obviously contain contributions of both the central transition and the satellite transitions. The top trace of Fig.~\ref{fgr:KAlH4_27Al_centre} shows the third to fifth spinning sidebands of the same spectrum. These spinning sidebands contain only the well-resolved contributions of the inner and outer satellite transitions. The width of the symmetric spinning sidebands of the inner satellites is about 115 Hz ($FWHH$). From the positions of the spinning sidebands +5 to +3 and $-$3 to $-$5, we obtain $\delta_{cg}(\tfrac {3}{2}) = 107.7$\,ppm. With $\delta_{cg}(\tfrac {1}{2}) = 106.9$\,ppm we arrive at $\delta_{iso}=(107.6\pm 0.2)$\,ppm at $T=295$\,K. The margin of error given is a rather conservative estimate, taking into account possible errors in the calibration with the external standard and the above mentioned effect of a temperature increase by frictional heating under MAS.  

The well-resolved MAS NMR spectrum in Fig.~\ref{fgr:KAlH4_27Al_centre} does not only allow the correct isotropic chemical shift to be determined, but also the strength of the quadrupole coupling to be estimated. The splitting between the maxima of the spinning sidebands of the two pairs of satellite transitions amounts to $(2.3\pm0.1)$\,ppm. Eqn~(\ref{eq:a}) implies that the splitting is caused by the difference in the quadrupole-induced shifts. Eqn~(\ref{eq:z}) shows that for $I = \tfrac {5}{2}$ this difference is given by
\begin{equation}
  \delta_{qis}(\tfrac {5}{2}) - \delta_{qis}(\tfrac {3}{2}) = \frac{81}{4000} \frac{C_Q^2}{\nu_L^2}\left(1+\frac{\eta^2}{3}\right).
   \label{eq:c}
\end{equation}
Hence, the splitting observed in the spinning sidebands is directly related to the two parameters describing the quadrupole interaction: the quadrupole coupling constant $C_Q$ and the asymmetry parameter $\eta$. The quantity 
\begin{equation}
  P = C_Q\left(1+\frac{\eta^2}{3}\right)^{1/2}
  \label{eq:z3}
\end{equation}
is referred to as the quadrupolar interaction product\cite{Engelhardt99} or the second-order quadrupole effect (SOQE).\citep{Jakobsen89,Skibsted91} Since eqn~(\ref{eq:c}) yields a value of about 1.39\,MHz for $P$, the quadrupole coupling constant for the aluminium atom in \ce{KAlH4} must lie  somewhere in the range of 1.20\,MHz (if $\etaη = 1$) to 1.39\,MHz (if $\eta = 0$). Alternatively, the value of $P$ could be determined from the difference $\delta_{qis}(\tfrac {3}{2}) - \delta_{qis}(\tfrac {1}{2})$. However, this difference is only one third of that given in eqn~(\ref{eq:c}). Provided that the spinning sidebands of the outer satellite transitions are well resolved, calculations based on eqn~(\ref{eq:c}) are more precise.

We can even go one step further. At sufficiently low spinning speeds, the envelope of the spinning sidebands agrees with the lineshape one would obtain for a stationary sample.  The total spread of the resonance line caused by a pair of satellite transitions ($\Delta\nu_{TS}(m)$) depends on the quadrupole coupling constant $C_Q$, but is independent of the asymmetry parameter $\eta$.\cite{Taylor75} Hence, the range over which the spinning sidebands can be observed yields a rather good estimate for $C_Q$. If the range over which sidebands are actually detected is limited by the finite bandwidth of the probe, the excitation or the receiver, spectra with various frequency offsets can be taken to overcome these limitations. For the $^{27}$Al nucleus ($I = \tfrac {5}{2}$), the relation between $\Delta\nu_{TS}(m)$ and $C_Q$ reads as\cite{Taylor75,Freude00}  
\begin{equation}
  \Delta\nu_{TS}(\tfrac {5}{2})=2\Delta\nu_{TS}(\tfrac {3}{2})=\frac{3}{5}C_Q.
   \label{eq:d}
\end{equation}

\begin{figure}
  \includegraphics [width=10cm]{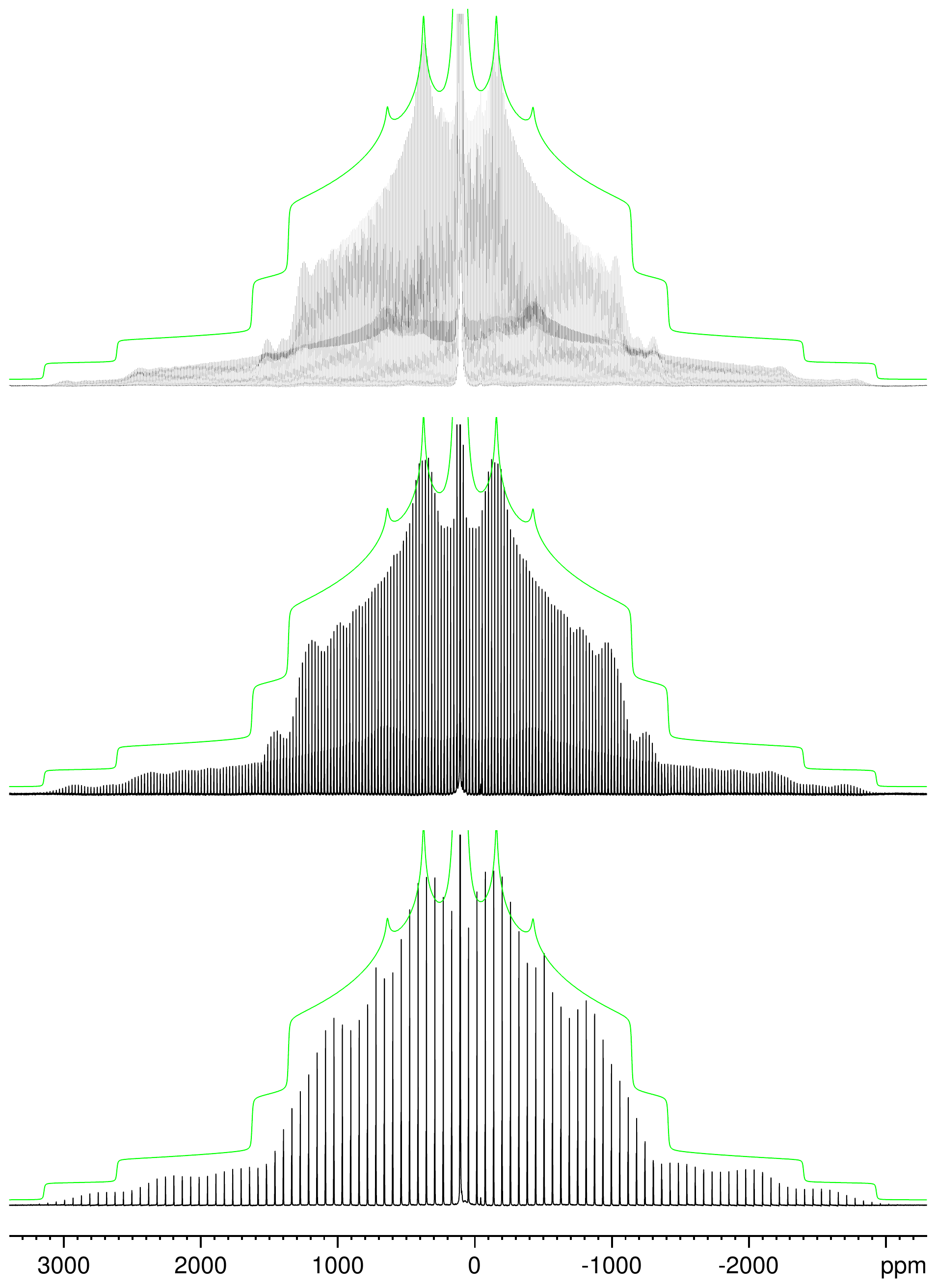}
  \caption{$^{27}$Al MAS NMR spectra of KAlH$_{4}$ measured at different spinning frequencies $\nu_{MAS}$: 1.4\,kHz (top), 3.0\,kHz (middle)and 8.0\,kHz (bottom). The intense lines of the central transition are cut off above the maxima of the inner satellite transitions, i.e., at about 2 to 7\% of their maximum height. The green lines depict the lineshape one would obtain for a non-spinning sample. For the simulation the parameters $\delta_{iso}=107.6$\,ppm, $C_Q=1.32$\,MHz and $\eta=0.65$ were used.}
  \label{fgr:KAlH4_27Al_full}
\end{figure}

$^{27}$Al MAS NMR spectra of KAlH$_{4}$ measured at different spinning frequencies are summarized in Fig.~\ref{fgr:KAlH4_27Al_full}. The spectrum in the middle shows the same spectrum as Fig.~\ref{fgr:KAlH4_27Al_centre}, but now in a wide range of 6700\,ppm corresponding to about 0.87\,MHz. The two sets of spinning sidebands caused by the inner and outer satellite transitions can easily be discerned. In the spectrum measured at the lowest spinning frequency (top trace in Fig.~\ref{fgr:KAlH4_27Al_full}), the sidebands of the inner satellites spread from 1620\,ppm to $-$1410\,ppm ($\Delta\nu_{TS}(\tfrac {3}{2})\approx395$\,kHz) and those of the outer satellites from 3130\,ppm to $-$2920\,ppm ($\Delta\nu_{TS}(\tfrac {5}{2})\approx788$\,kHz). Since the range of the spinning sidebands of the outer satellite is about twice as wide as the range of the inner satellite, we can exclude detection problems caused by a limited bandwidth. With the estimate for $\Delta\nu_{TS}(\tfrac {3}{2})$,  eqn~(\ref{eq:d}) yields a value for the quadrupole coupling constant of $C_Q\approx1.32$\,MHz. This result falls into the range derived above from the splitting of the spinning sidebands of the satellite transitions. From this value for $C_Q$ and the above determined value for the quadrupolar interaction product $P$, we obtain a first estimate for the assymetry parameter: $\eta\approx0.6$. The non-vanishing asymmetry parameter corresponds to the fact that the aluminium atoms are located in Wyckoff position $4c$ in \textit{Pnma} \ce{KAlH4}. There is no $C_n$ axis with $n\geq3$ going through this position.

A more precise estimate of the asymmetry parameter can be obtained when the ratio between the splitting of the maxima of the inner satellite transition ($\Delta\nu_{M}(\tfrac {3}{2})$) and the total spread of this transition ($\Delta\nu_{TS}(\tfrac {3}{2})$) is used. From the definition of $\eta$ (eqn~(\ref{eq:z2})) and the fact that the field gradient tensor is traceless, it follows that for any transition $m$ the asymmetry parameter is related to the characteristic points of the first-order quadrupole lineshape by the following equation:
\begin{equation}
  \eta=1-\frac{2\Delta\nu_M(m)}{\Delta\nu_{TS}(m)}.
   \label{eq:e}
\end{equation}
The spectrum measured at $\nu_{MAS} = 1.4$\,kHz (top trace in Fig.~\ref{fgr:KAlH4_27Al_full}) yields $\Delta\nu_M(\tfrac {3}{2})\approx69$\,kHz. With the above derived value for $\Delta\nu_{TS}(\tfrac {3}{2})$ we obtain $\eta\approx0.65$. This value of the asymmetry parameter and the above derived one for the quadrupole coupling constant have been used to simulate the lineshape  of the satellite transitions one would obtain for a non-spinning sample (\textit{cf.} Fig.~\ref{fgr:KAlH4_27Al_full}).

\begin{figure}
 \includegraphics [width=8.3cm]{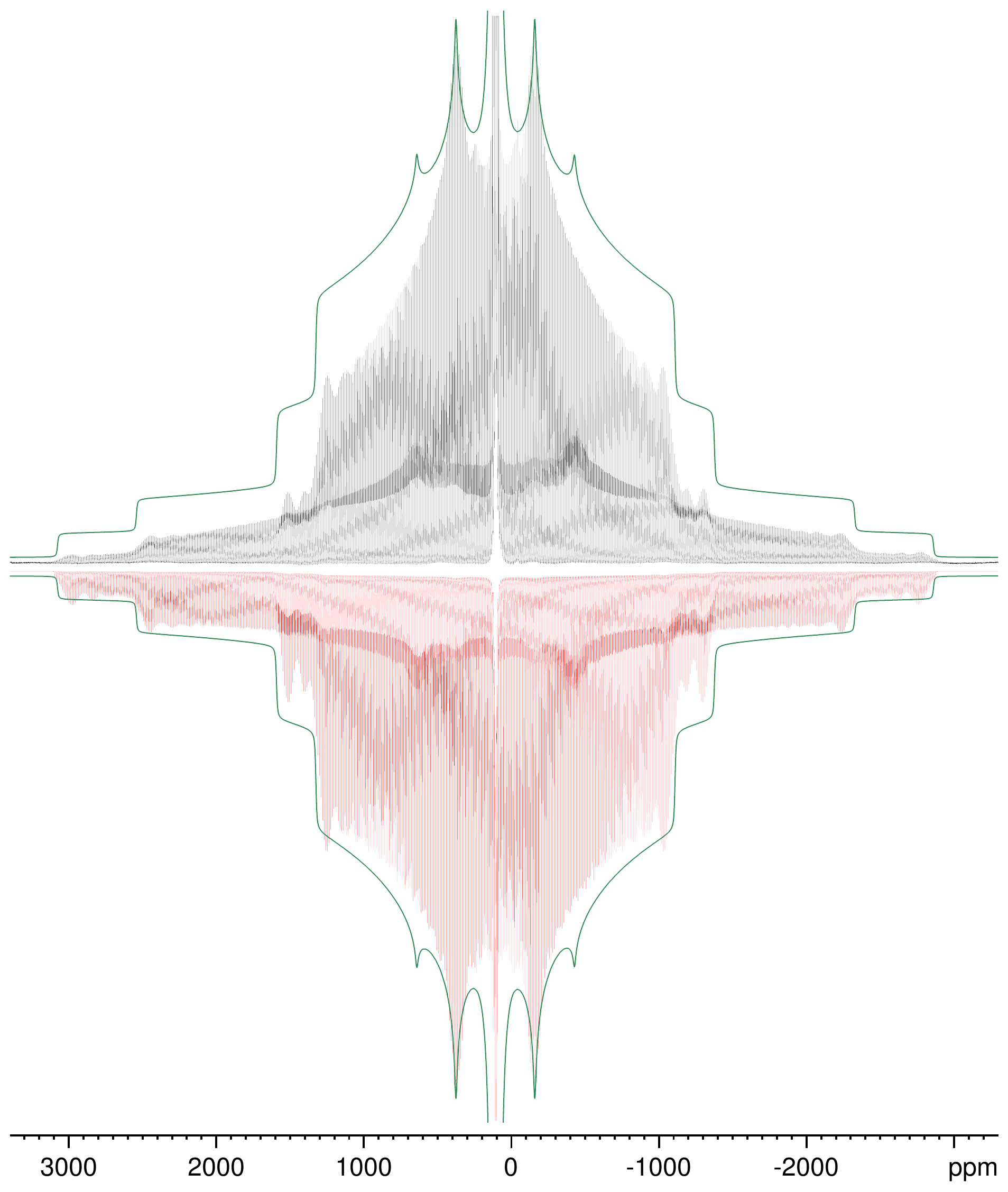}
  \caption{Experimental $^{27}$Al~MAS~NMR spectrum of \ce{KAlH4} measured at  $\nu_{MAS} = 1.4$\,kHz (black) and its simulation (red) with the parameters $\delta_{iso}=107.6$\,ppm, $C_Q = 1.29$\,MHz and $\eta = 0.64$. To ease the comparison, the simulated MAS~NMR~spectrum has been inverted. Again, the green lines depict the lineshape one would obtain for a non-spinning sample. }
   \label{fgr:KAlH4_27Al_simulation}
\end{figure}

Fig.~\ref{fgr:KAlH4_27Al_full} shows that spinning at 1.4\,kHz is still too fast to avoid significant deviations of the envelope of the sideband pattern from the lineshape of the stationary sample. Almost three decades ago, Jakobsen and co-workers have demonstrated that these characteristic intensity modulations of the spinning sidebands of the satellite transitions can be used to determine the values of the parameters $C_Q$ and $\eta$ with high accuracy.\cite{Jakobsen89,Skibsted91} Nowadays, the software to perform this kind of lineshape simulations comes with the spectrometer. Using the above derived values for the parameters $\delta_{iso}$, $C_Q$ and  $\eta$ as starting values for the fit, only a few iteration cycles are necessary to reach a nice agreement between experiment and simulation, as shown in Fig.~\ref{fgr:KAlH4_27Al_simulation}. The reduced intensities in the outer wings of the experimental sideband pattern are mainly caused by the limited excitation width of the rf pulses used. The quality of the fit can better be judged from the enlarged version of Fig.~\ref{fgr:KAlH4_27Al_simulation} given as Fig.~S2 in the ESI.

By varying the simulation parameters and carefully comparing the simulated with the experimental spectrum we finally derived the following values for the parameters describing the quadrupole interaction: $C_Q = (1.29\pm0.02)$\,MHz and $\eta = 0.64\pm0.02$. A comparison between experiment and simulation for $\nu_{MAS} = 3.0$\,kHz (Fig.~S3) and a complete list of the parameters used for the simulation of the sideband patterns (Table S1) can be found in the ESI.

From the strength of the quadrupole coupling found for $^{27}$Al in \ce{KAlH4}, it is apparent why the centreband of the central transition in Fig.~\ref{fgr:KAlH4_27Al_centre} does not show any indication of second-order quadrupolar broadening. The mathematical expressions for the characteristic features of the lineshape for a quadrupolar nucleus with a half-integer spin $I$ under MAS conditions were derived decades ago. With the expressions given by Freude and Haase \cite{FreudeHaase} for the minimum and maximum frequencies, the total width of the MAS  centreband of the central transition is obtained as
\begin{equation}
  \Delta\nu_{TS,MAS}^{(2)}(\tfrac {1}{2})=\frac{9}{56}\frac{I(I+1)-\frac{3}{4}}{I^2(2I-1)^2}\frac{C_Q^2}{\nu_L}\left(1+\frac{\eta}{6}\right)^2.
   \label{eq:f}
\end{equation}
For $I = \tfrac {5}{2}$ and the above determined values of $C_Q$ and $\eta$, eqn~(\ref{eq:f}) yields  $\Delta\nu_{TS,MAS}(\tfrac {1}{2})= 201$\,Hz. The total broadening caused by second-order quadrupole interaction exceeds the experimental $FWHH$ of the centreband of the central transition by only a few Hertz. Other broadening mechanisms such as the homonuclear dipole-dipole interaction or the line broadening caused by structural defects mask the effect of second-order quadrupole interaction.

The two-dimensional one pulse (TOP) presentation  of 1D~MAS~NMR spectra \cite{Massiot06}can offer an alternative convenient way to extract the relevant parameters for half-integer quadrupolar nuclei. We have added such a presentation of an experimental spectrum shown in Fig.~\ref{fgr:KAlH4_27Al_full} as Fig.~S4 to the ESI.\dag~  The line positions of the central transition and the satellite transitions can be read out directly from this plot. The lack of any significant second-order quadrupolar broadening is obvious. 

Provided that neither other broadening mechanisms nor the presence of by-products obscure the characteristic features of the quadrupolar lineshape, the parameters describing the quadrupole interaction can alternatively be derived from spectra measured without spinning the sample under the magic angle. Because of the width of these spectra, corresponding to a very fast decay of the NMR signal, the spectra of stationary samples have to be measured with an appropriate echo technique to overcome the inevitable dead time problems after a strong radio pulse. For molecular aluminophosphate compounds containing aluminium nuclei with quadrupole coupling constants slightly larger than the one observed here, the Solomon echo sequence \cite{Solomon58} has successfully been applied as an alternative approach for the determination of $C_Q$ and $\eta$. \cite{Azais02}

The values determined for  $C_Q$ and $\eta$ in the present paper do not agree with any of the pairs of quadrupole coupling parameters determined by Tarasov \textit{et al.}\cite{Tarasov00} for the three different sites they found ($C_Q$ in MHz/$\eta$: 0.58/0.1, 2.07/0.0, 2.8/0.05). The authors inferred the presence of these different sites with different quadrupole couplings from a pattern of very weak resonance lines on  both sides of the strong central line of \ce{KAlH4}. On the other hand, the authors mention the presence of a by-product giving rise to a broad line in the $^{27}$Al NMR spectrum. Unfortunately, no diffraction data are  reported. Based on chemical analysis, the authors state a phase purity of no more than 95--97\%. Hence, we refrain from speculating about the origin of this obvious disagreement with our data.

 Sorte \textit{et al.}\cite{Sorte14} reported first-order quadrupole patterns "from both MAS and quadrupole echoes, with no sharp singularities." The lack of well-defined lineshapes is obviously due to the rather questionable quality of the \ce{KAlH4} sample used. It should be noted that the NMR measurements were performed on the unannealed material, i.e., not on the material whose XRD pattern is given in the paper. Apart from the obvious presence of a significant amount of metallic aluminium, the authors mention several reflections in the XRD pattern of the unannealed material that do not correspond to \textit{Pnma}  \ce{KAlH4} and assign them to unknown impurities or to other crystal phases of \ce{KAlH4}.\cite{Sorte14} With these characteristics of the material it is debatable as to which extent the experimental results obtained mirror  intrinsic properties of  \ce{KAlH4} and not the peculiarities of the sample measured. 
 
Ares \textit{et al.}\cite{Ares16} reported rather well resolved  $^{27}$Al MAS NMR spectra for  \ce{KAlH4} samples synthesized via direct hydrogenation under mechanochemical  conditions with and without the presence of different metal chlorides. To further compare our material with those used in this recent study, we measured  $^1$H MAS~NMR spectra at different spinning frequencies (\textit{cf.} Fig.~S5 in the ESI\dag). The $^1$H resonance line of \ce{KAlH4} at about 3\,ppm ($FWHH\approx1.3$\,kHz at $\nu_{MAS}$ = 8\,kHz) is very similar to those reported by Ares \textit{et al.},\cite{Ares16} except that our spectra do not show any indication of the presence of KH. Their $^{27}$Al MAS NMR spectra are also very similar to those we recorded for samples directly after the salt metathesis, i.e.,  prior to the recrystallisation procedure described above. We believe that the lack of sharp features in the first-order quadrupolar spinning-sideband patterns noticed by the authors\cite{Ares16} is typical of materials produced by ball milling since they usually consist of very small particles with a large number of defects. In particular, the presence of strong spinning sidebands of the central transition masks the minimum in the envelope of the spinning sidebands of the inner satellite transition in the centre of the line. This leads to the erroneous assumption  of a small or even vanishing asymmetry parameter. Furthermore, we have never achieved a sufficient resolution of the sidebands of the outer and inner satellite transition for samples obtained directly from ball milling. The attempt undertaken by  Ares \textit{et al.}\cite{Ares16} to fit the sideband envelope by using the parameters reported by Tarasov \textit{et al.}\cite{Tarasov00} was bound to fail. Hence, we are left without any $^{27}$Al NMR study on \ce{KAlH4} we can compare our results with.

For the much better studied \ce{NaAlH4}, which crystallises in the tetragonal space group ${I4_{1}/a}$,\cite{Hauback03} a significantly stronger quadrupole coupling has been found ($C_Q = 3.15$\,MHz\cite{Zhang09}). Since the Al atoms in \ce{NaAlH4} are located in Wyckoff  position $4b$, i.e., on a $C_4$ axis, the asymmetry parameter is zero. For the monoclinic \ce{LiAlH4} (space group $P2_{1}/c$\cite{Hauback02}) an even stronger quadrupole coupling ($C_Q = 3.90$\,MHz\cite{Kellberg90, Wiench04}) has been reported. The asymmetry parameter for \ce{LiAlH4} has been determined as 0.30\cite{Kellberg90} or 0.24.\cite{Wiench04} For both the orthorhombic and the tetragonal phases of \ce{CsAlH4},  centrebands of the $^{27}$Al central transition have been observed \cite{Krech14} that are almost as narrow and without any indication of second-order quadrupole interaction as the centrebands obtained here for  \ce{KAlH4}. Hence, we estimate that the strength of the quadrupole coupling for both modifications of \ce{CsAlH4} is similar to that of \ce{KAlH4}.

\subsection*{$^{39}$K MAS NMR spectroscopy}

Because of the relatively low gyromagnetic ratio of the $^{39}$K isotope (natural abundance: 93.3\%, $I = \tfrac {3}{2}$), the intensity of the resonance line is much weaker than in the case of the same number of $^{27}$Al nuclei at the same field. Furthermore, quadrupole coupling constants of up to 3.2\,MHz have been found for the cation in simple potassium salts. \cite{Moudrakovski07} For $I=\frac{3}{2}$  a quadrupole coupling of this strength leads to much wider resonance lines than in the case of  $I=\frac{5}{2}$. As can be seen in eqn~(\ref{eq:f}), the total width of the central transition due to second-order quadrupole interaction scales with the factor $(I(I+1)-\frac{3}{4})(I^{2}(2I-1)^2)^{-1}$. However, with the sensitivity of modern NMR spectrometers operating at sufficiently high magnetic fields the detection of $^{39}$K NMR lines for solids is usually no serious problem, provided that the samples are well defined. 

\begin{figure}
  \includegraphics [width=10cm]{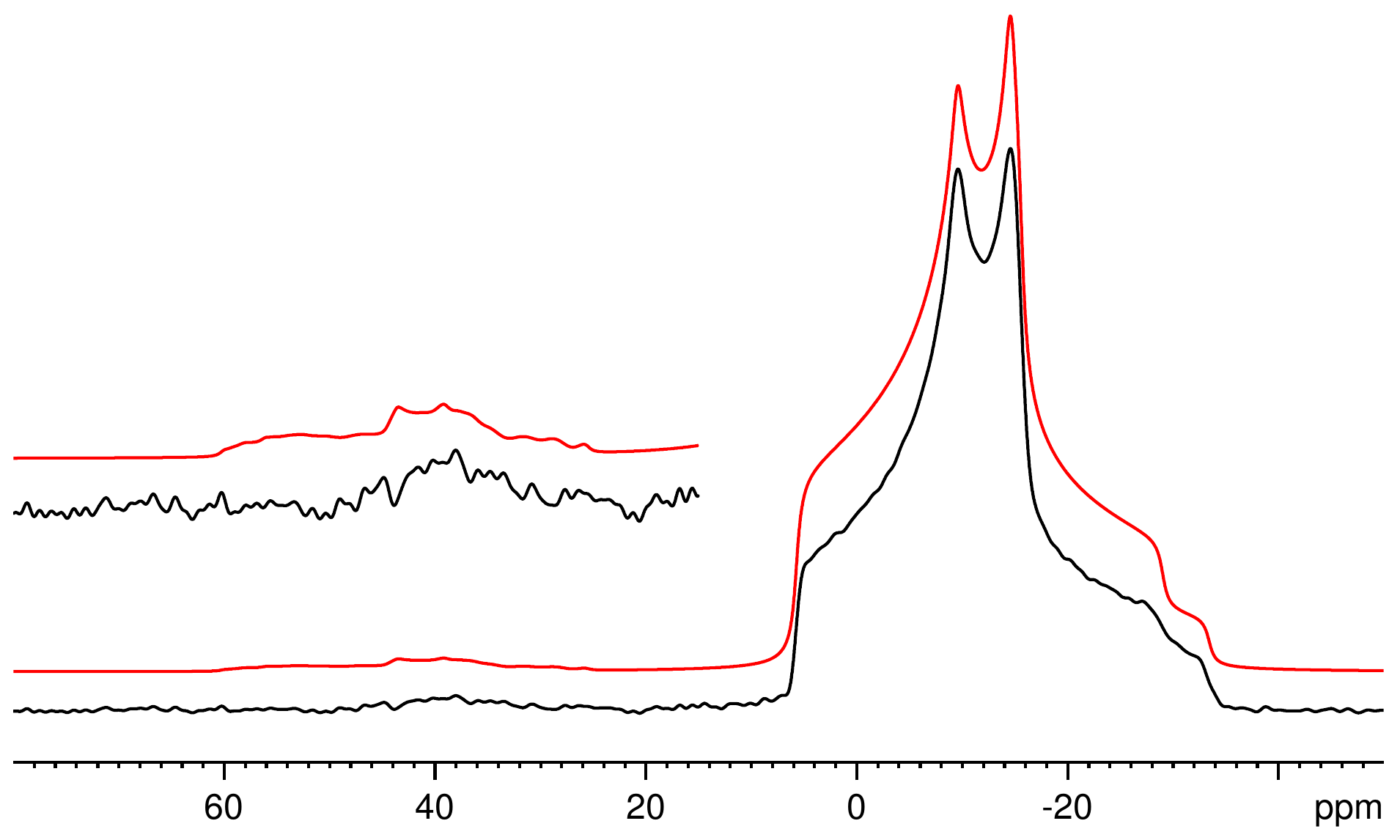}
  \caption{Experimental $^{39}$K~MAS~NMR spectrum of \ce{KAlH4} measured at  $\nu_{MAS} = 6 $\,kHz (black). The inset shows the region of the centreband of the satellite transitions enlarged by a factor of four. The red lines depict the theoretical lineshapes of the centrebands of both the central transition and the satellite transitions simulated with the parameters $\delta_{iso}=6.1$\,ppm, $C_Q = 0.562$\,MHz and $\eta = 0.74$. }
  \label{fgr:KAlH4_39K}
\end{figure}

A $^{39}$K~MAS~NMR spectrum of our \ce{KAlH4} sample is shown in Fig.~\ref{fgr:KAlH4_39K}. The shape of the centreband of the central transition is clearly governed by second-order quadrupole interaction. The spectrum exhibits the discontinuities characteristic of this type of interaction. The overall lineshape indicates a non-vanishing asymmetry parameter.\cite{Freude00} Since the potassium cation sits in Wyckoff position $4c$ in \textit{Pnma} \ce{KAlH4},\cite{Hauback05} this finding is not surprising. The total spread of the resonance line $\Delta\nu_{TS,MAS}(\tfrac {1}{2})$ is about 920\,Hz. Assuming an asymmetry parameter of 0.8, eqn~(\ref{eq:f}) yields an estimate for the quadrupole coupling constant: $C_Q \approx 0.5$\,MHz. These values were used as starting values for an iterative simulation of the lineshape.  As Fig.~\ref{fgr:KAlH4_39K} shows, the simulation does not only reproduce the characteristic features of the centreband of the central transition, but identifies the spectral intensity on the low-field side of this line as the much weaker centreband of the satellite transitions $(-\tfrac{3}{2}\leftrightarrow-\tfrac{1}{2})$ and $(+\tfrac{1}{2}\leftrightarrow+\tfrac{3}{2})$. There is no indication of the presence of any potassium-containing by-product. By comparing simulated and experimental spectra measured at different spinning speeds we derived the following parameters: $\delta_{iso}=(6.1\pm0.2)$\,ppm, $C_Q = (0.562\pm0.005)$\,MHz and $\eta = 0.74\pm0.02$. 

At first glance, there is a large discrepancy between our data and those reported by Tarasov \textit{et al.}: $\delta_{iso}=(-49\pm5)$\,ppm, $C_Q = (1.12\pm0.07)$\,MHz and $\eta = (0.83\pm0.05)$.\cite{Tarasov00} However, this discrepancy can easily be resolved. Tarasov \textit{et al.}\cite{Tarasov00} used a procedure outlined by Granger \cite{Granger89} to extract the parameters of the quadrupole interaction from their spectrum measured for a stationary sample.  Unfortunately, they overlooked an obvious typo in the original paper. The correct equation for $C_Q$ should read
\begin{equation}
  C_Q=\frac{2}{3}I(2I-1)\left(\frac{\nu_{L}a}{I(I+1)-\frac{3}{4}}\right)^{1/2},
  \label{eq:z4}
\end{equation}
with $a= 144\Delta\nu_{TS}/(25+22\eta+\eta^2)$. For $I = \tfrac {3}{2}$, using a factor $(2I+1)$ instead of the correct $(2I-1)$ accounts for a factor of two for the quadrupole coupling constant. Hence, our data for the quadrupole coupling actually agree very nicely with those found by Tarasov \textit{et al.}\cite{Tarasov00} Eqn~(\ref{eq:z4}) follows directly from the mathematical expressions found for the minimum and maximum frequencies of the central transition for a half-integer nucleus that is subjected to second-order quadrupole interaction.\citep{FreudeHaase} In analogy to eqn~(\ref{eq:f}) for a sample under MAS conditions, the total spread of the resonance line for a stationary sample is given by: 
\begin{equation}
  \Delta\nu_{TS}^{(2)}(\tfrac {1}{2})=\frac{1}{64}\frac{I(I+1)-\frac{3}{4}}{I^2(2I-1)^2}\frac{C_Q^2}{\nu_L}(25+22\eta+\eta^2).
   \label{eq:g}
\end{equation}

The discrepancy concerning the  chemical shift data can be explained even more easily. What Tarasov \textit{et al.}\cite{Tarasov00} refer to as  isotropic chemical shift is not $\delta_{iso}$, but the shift of the centre of gravity of the resonance line caused by second-order quadrupole interaction, i.e., the quadrupole-induced shift $\delta_{qis}$.
From eqn~(\ref{eq:z}) it follows that for $I = \tfrac {3}{2}$ this shift is given by
\begin{equation}
  \delta_{qis}(\tfrac {1}{2}) = -\frac{1}{40} \frac{C_Q^2}{\nu_L^2}\left(1+\frac{\eta^2}{3}\right).
   \label{eq:z5}
\end{equation}
With our above determined parameters of the quadrupole coupling, eqn~(\ref{eq:z5}) yields a value of $-47.7$\,ppm for $\delta_{qis}$ at the Larmor frequency of $\nu_{L}=14.0$\,MHz used by Tarasov \textit{et al.}\cite{Tarasov00} Hence, our data agree very well with those of the earlier study. 

In the more recent paper by Sorte \textit{et al.},\cite{Sorte14} a $^{39}$K NMR line at $-10.9$\,ppm with $FWHH=800$\,Hz and a total spread of about 3\,kHz is reported for a stationary sample. Under the same experimental conditions, Tarasov \textit{et al.}\cite{Tarasov00} observed  a resonance line with a splitting of 614\,Hz and a total spread   $\Delta\nu_{TS}^{(2)}(\tfrac {1}{2})= 5.1$\,kHz. According to eqn~(\ref{eq:g}), one would expect a total spread of the resonance line of about 4.92\,kHz from the above determined parameters. Hence, we have to conclude that the $^{39}$K NMR line reported by Sorte \textit{et al.}\cite{Sorte14} is not caused by the potassium ions in \textit{Pnma} \ce{KAlH4}.

\section*{Conclusions}
We have shown that potassium aluminium tetrahydride \ce{KAlH4} of high phase purity can easily be synthesized via a mechanochemical route. The thus obtained material crystallises in space group \textit{Pnma} (62). We believe that a sufficient quality of the hydride sample is not only important for obtaining a convincing resolution in the MAS NMR spectra, but will be of paramount importance in future studies of dynamical properties and the decomposition process.

Using a well-established procedure and spectral simulation we have derived precise data for the isotropic chemical shift and the quadrupole coupling for both nuclei studied. Apart from the quadrupole coupling data for $^{27}$Al, the results agree fairly well with those of an earlier study by Tarasov \textit{et al.}\cite{Tarasov00}

Provided that samples of sufficient crystallinity and phase purity are used, the approach applied in the present paper should also work for other alkali aluminium hydrides and, more generally, for many other materials containing half-integer nuclei experiencing small to medium-sized quadrupole couplings.

\section*{Conflicts of interest}
There are no conflicts to declare.

\section*{Acknowledgement}

We are very much indebted to Dr Thomas Bernert for measuring and evaluating the X-ray data.

\subsection*{Electronic Supplementary Information (ESI)}
    Additional $^{27}$Al MAS NMR spectra, a table with the parameters used for the simulation of the $^{27}$Al MAS NMR sideband patterns and $^{1}$H MAS NMR spectra are given in the ESI.

\bibliography{KAlH4_arxiv}

\end{document}